\documentstyle[sprocl,epsfig]{article}

\newcommand{\dgs}{d^{\gamma}{\!\scriptstyle (}s{\scriptstyle )}\,} 
\newcommand{\dzs}{d^{Z}{\!\scriptstyle (}s{\scriptstyle )}\,} 
\newcommand{\mIm}{\,\mbox{\small $\Im$m\,}} 
\newcommand{\eRe}{\,\mbox{\small $\Re$e\,}} 
\newcommand\sfrac[2]{{\textstyle \frac{#1}{#2}}}
\newcommand{\Pol}{\mbox{\small P}}
\newcommand{\be}{\begin{eqnarray}} 
\newcommand{\ee}{\end{eqnarray}} 
\newcommand{\nee}{\nonumber \end{eqnarray}}

\newcommand{\CPbar}
  {\begin{picture}(14,7)
   \put(0,0){CP} \thicklines \put(0,0){\line(2,1){14}}
   \end{picture}$\:$}

\begin{document}

\vspace*{-45mm}
\begin{flushright}
  hep--ph/9712380 \\
  HEPHY--PUB 681/97 \\
  INRNE--TH--97/14 \\
  UWThPh--1997--48 
\end{flushright}
\vspace*{6mm}

\title{SUSY induced CP violation in \mbox{$e^{+}e^{-} \to t\,\bar{t}$}}

\author{
A. Bartl${}^{\mathbf{1}}$, 
E. Christova${}^{\mathbf{2}}$, 
T. Gajdosik${}^{\mathbf{3}\dagger}$,
W. Majerotto${}^{\mathbf{3}}$ }

\address{
${}^{\mathbf{1}}$Institut f\"ur Theoretische Physik, Universit\"at Wien, 
Vienna, Austria; \\
${}^{\mathbf{2}}$Institute of Nuclear Research and Nuclear Energy, 
  Sofia, Bulgaria; \\
${}^{\mathbf{3}}$Institut f\"ur Hochenergiephysik, 
  \"Osterreichische Akademie der Wissenschaften, Vienna, Austria;
} 


\maketitle\abstracts{ 
The CP violating electric and weak dipole moment form 
factors of the top quark, $\dgs$ and $\dzs$, appear in the 
process \mbox{$e^{+}e^{-} \to t\,\bar{t}$}. We present a 
complete analysis of these dipole moment formfactors within the 
Minimal Supersymmetric Standard Model with complex parameters. We 
include gluino, chargino, and neutralino exchange in the loop of the 
\mbox{$\gamma t\bar{t}$} and \mbox{$Z t\bar{t}$} vertex.
We give analytic expressions and present numerical results 
for the asymmetries that measure these formfactors.}

\footnotetext{$\dagger$ Talk presented at the 
\emph{International Workshop ``Beyond the Standard Model: 
from Theory to Experiment}, 
October 13 -- 17, 1997, Valencia, Spain.}

\section{Introduction}
The large mass of the top quark 
allows one to probe physics at a high energy scale, where 
new physics might show up. In the last years a number of 
papers~\cite{{bcgm:1},{we}} considered \CPbar observables in top 
quark production (and decays) as tests for new physics. In 
$e^{+} e^{-}$ annihilation these effects are due to the weak and 
electric dipole moment formfactors $\dzs$ and $\dgs$ of the top quark.
In general, the $\gamma t \bar{t}$ and $Z t\bar{t}$ vertices 
${\cal V}_{\gamma}^{t}$ and ${\cal V}_{Z}^{t}$ 
including the \CPbar formfactors are 
\be
  e ( {\cal V}_{\gamma}^{t} )_{\mu} 
&=& e ( \sfrac{2}{3} \gamma_{\mu} 
    - i \dgs ({\cal P}_{\mu}/m_{t}) \gamma_{5} )
\\
  g_{Z} ( {\cal V}_{Z}^{t} )_{\mu} 
&=& g_{Z} ( \gamma_{\mu} ( g_{V} + g_{A} \gamma_{5} ) 
    - i \dzs ({\cal P}_{\mu}/m_{t}) \gamma_{5} ) \: ,
\ee
where \mbox{${\cal P}_{\mu} = p_{t\,\mu} - p_{\bar{t}\,\mu}$}, 
$g_{V} = (1/2) - (4/3) \sin^{2}\Theta_{W}$,
$g_{A} = - (1/2)$, 
$g_{Z} = g / (2 \cos \Theta_{W})$, 
and 
\mbox{$g = e/\sin\Theta_{\scriptscriptstyle W}$} with $e$ the 
electromagnetic coupling constant and 
$\Theta_{\scriptscriptstyle W}$ the Weinberg angle.
 
In the Standard Model (SM) \CPbar can 
appear only through the phase in the CKM--matrix. 
The dipole moment formfactors $\dgs$ and $\dzs$ for quarks 
are at least two--loop order effects and hence very small. 
In the Minimal Supersymmetric Standard Model (MSSM)~\cite{Kane} 
with complex parameters additional complex couplings are possible
leading to \CPbar within one generation 
at one--loop level~\cite{Dugan}. If the masses of SUSY particles 
are not very much higher than the mass of the top, one expects 
SUSY radiative corrections to induce non--neglible contributions to 
$\dgs$ and $\dzs$. They are calculated in~\cite{{we},{bcgm}}. Although 
the gluino contribution is proportional to the strong coupling 
$\alpha_{s}$ the chargino contribution, which is proportional to 
$\alpha_{w}$\mbox{$( = g^{2}/(4\pi)$)} 
can be equally important (see also~\cite{Osh1}). This is due to 
threshold enhancements and the large Yukawa couplings: 
\mbox{$Y_{t} = m_{t}/(\sqrt{2} m_{\scriptscriptstyle W} \sin\beta)$} and 
\mbox{$Y_{b} = m_{b}/(\sqrt{2} m_{\scriptscriptstyle W} \cos\beta)$}. 
In general the neutralino contribution turns out to be smaller. 
However, there are cases where it is important. 

\section{Complex couplings in the MSSM}
In the MSSM the higgsino mass parameter $\mu$ and the trilinear 
scalar coupling parameters $A_{t}$ and $A_{b}$ can be complex and 
thus provide \CPbar phases. Assuming unifivation there are 
constraints~\cite{{Osh1},{Garisto}} 
on the phases from the measurement of the electric dipole moment 
(EDM) of the neutron. Being less restrictive we only 
assume unification of the gauge couplings: 
$m_{\tilde{g}} = (\alpha_{s}/\alpha_{2}) M \approx 3 M$ and 
$M^{\prime} = \sfrac{5}{3} \tan^{2}\Theta_{\scriptscriptstyle W} M$.
The calculation of the dipole moment formfactors requires the 
diagonalization of the squark, chargino, and neutralino mass matrices.
We use the singular value decomposition~\cite{SingW} to diagonalize 
the complex neutralino and chargino mass matrices. 

The gluino contribution has been calculated in \cite{we}. The 
chargino and neutralino contributions~\cite{bcgm} 
depend on the gaugino and higgsino couplings, as well as on 
the squark mixing angles and phases. Quite generally, the terms 
proportional to $Y_{t}$ are important. We have also included the terms 
proportional to the bottom Yukawa coupling $Y_{b}$ which cannot be 
neglected for large $\tan\beta$. 

The loop integrations for $\dgs$ and $\dzs$ are done with 
the Passarino--Veltman three point functions 
$C_{0}$, $C_{i}$, and $C_{ii}$ ($i = 1,2$)~\cite{PaVe}. 

More details about the calculation of 
$\dgs$ and $\dzs$ can be found in~\cite{bcgm}. 

\section{Asymmetries}
With the formfactors $\dgs$ and $\dzs$ we can define asymmetries that 
measure the $t$--polarization~\cite{cd}. The first 
asymmetry we consider is the \emph{energy asymmetry} in $b$ or $\bar{b}$ 
\begin{equation} 
\mbox{A}^{\! E}_{b(\bar{b})} (s)
= 
\sfrac{ \#_{b(\bar{b})}(E_{b(\bar{b})}>E_{0}) 
      - \#_{b(\bar{b})}(E_{b(\bar{b})}<E_{0})}
      { \#_{b(\bar{b})}(E_{b(\bar{b})}>E_{0}) 
      + \#_{b(\bar{b})}(E_{b(\bar{b})}<E_{0})}
\: ,
\end{equation}
where $E_{0} = \sqrt{s} ( m_{t}^{2} - m_{\scriptscriptstyle W}^{2} )
/(4 m_{t}^{2})$ is the average energy of the $b$ or $\bar{b}$. 
The corresponding \CPbar quantity (Figure~1) is 
\begin{equation} 
\mbox{R}(s) = \mbox{A}^{\! E}_{b} (s) - \mbox{A}^{\! E}_{\bar{b}} (s)
= - 8 \, \alpha_{b} \, \beta_{t} \, \mIm[ H_{1} ] / N_{t}
\end{equation}
where $N_{t}$ is a normalization factor proportional to the total cross 
section, $\beta = \sqrt{1 - 4 m_{t}^{2}/s}$, $\alpha_{b} = 
( m_{t}^{2} - 2 m_{\scriptscriptstyle W}^{2} ) / 
( m_{t}^{2} + 2 m_{\scriptscriptstyle W}^{2} )$,
\be 
H_{1} 
&=& ( 1 - \Pol \bar{\Pol} ) H_{1}^{s} 
+ ( \Pol - \bar{\Pol} ) H_{1}^{a}
\: ,
\\
H_{1}^{s} 
&=& ( \sfrac{2}{3} - c_{V} g_{V} h_{Z} ) \dgs 
- ( \sfrac{2}{3} c_{V} h_{Z} 
- ( c_{V}^{2} + c_{A}^{2} ) g_{V} h_{Z}^{2}
) \dzs 
\: ,
\\
H_{1}^{a}
&=& - c_{A} g_{V} h_{Z} \dgs
- ( \sfrac{2}{3} c_{A} h_{Z} 
  - 2 c_{V} c_{A} g_{V} h_{Z}^{2} ) \dzs 
\: ,
\ee
$\Pol$($\bar{\Pol}$) are the $e^{-}$($e^{+}$) longitudinal polarizations, 
$c_{V} = - (1/2) + 2 \sin^{2}\Theta_{\scriptscriptstyle W}$ and 
$c_{A} = (1/2)$ are the \mbox{$Z e^{+}e^{-}$} couplings, and $h_{Z} = 
s / [ ( s - m_{Z}^{2} ) \sin^{2} 2 \Theta_{\scriptscriptstyle W} ]$.
\begin{figure*}
$\begin{array}{ccc}
\begin{minipage}[l]{158pt} 
\begin{footnotesize}
\begin{picture}(158,155)(-2,0)
\put( 3,   1){\mbox{\epsfig{file=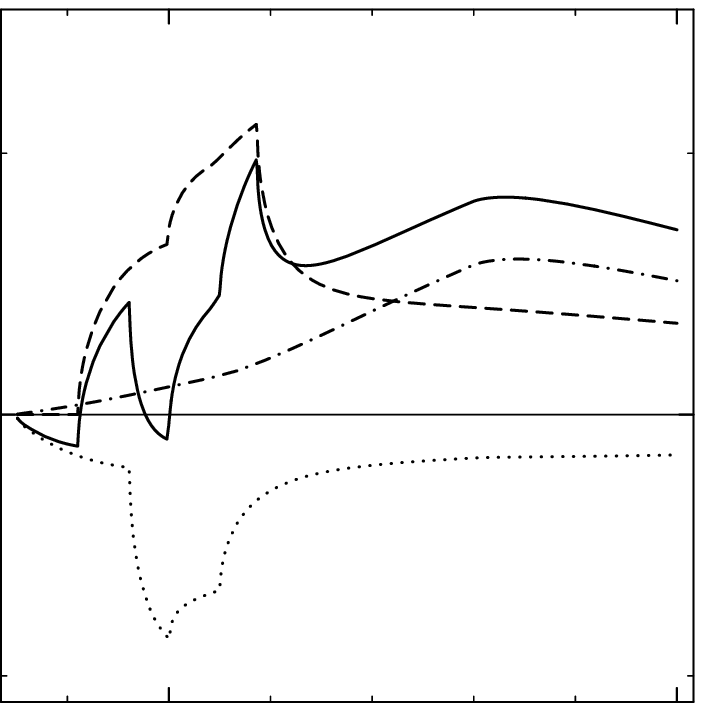,width=150pt}}}
\put( 8, 145){\makebox(0,0)[lt]{$\mbox{R}(s) \, \times 10^{4}$}}
\put(  0,148){\makebox(0,0)[ r]{ 3}}
\put(  0,120){\makebox(0,0)[ r]{ 2}}
\put(  0, 92){\makebox(0,0)[ r]{ 1}}
\put(  0, 64){\makebox(0,0)[ r]{ 0}}
\put(  0, 36){\makebox(0,0)[ r]{-1}}
\put(  0,  8){\makebox(0,0)[ r]{-2}}
\put( 15, -2){\makebox(0,0)[t]{400}}
\put( 37, -2){\makebox(0,0)[t]{500}}
\put( 59, -2){\makebox(0,0)[t]{600}}
\put( 81, -2){\makebox(0,0)[t]{700}}
\put(103, -2){\makebox(0,0)[t]{800}}
\put(125, -2){\makebox(0,0)[t]{900}}
\put(147,  7){\makebox(0,0)[br]{$\sqrt{s}/$GeV}}
\end{picture}
\caption{R$(s)$ (full line); contributions: 
chargino (dashed line), 
neutralino (dotted line), 
gluino (dashed--dotted line).}
\end{footnotesize}
\end{minipage} 
&\hspace{10pt}&
\begin{minipage}[l]{158pt} 
\begin{footnotesize}
\begin{picture}(158,155)(-2,0)
\put( 3,   1){\mbox{\epsfig{file=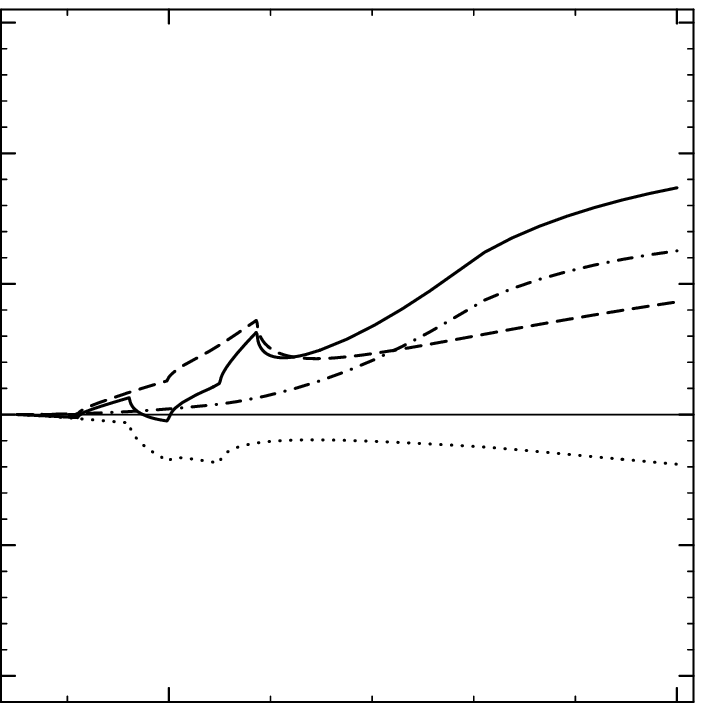,width=150pt}}}
\put( 8, 145){\makebox(0,0)[lt]{${\mathcal O}^{FB}(s) \, \times 10^{3}$}}
\put(  0,148){\makebox(0,0)[ r]{ 3}}
\put(  0,120){\makebox(0,0)[ r]{ 2}}
\put(  0, 92){\makebox(0,0)[ r]{ 1}}
\put(  0, 64){\makebox(0,0)[ r]{ 0}}
\put(  0, 36){\makebox(0,0)[ r]{-1}}
\put(  0,  8){\makebox(0,0)[ r]{-2}}
\put( 15, -2){\makebox(0,0)[t]{400}}
\put( 37, -2){\makebox(0,0)[t]{500}}
\put( 59, -2){\makebox(0,0)[t]{600}}
\put( 81, -2){\makebox(0,0)[t]{700}}
\put(103, -2){\makebox(0,0)[t]{800}}
\put(125, -2){\makebox(0,0)[t]{900}}
\put(147,  7){\makebox(0,0)[br]{$\sqrt{s}/$GeV}}
\end{picture}
\caption{${\mathcal O}^{FB}(s)$ (full line); contributions: 
chargino (dashed line), 
neutralino (dotted line), 
gluino (dashed--dotted line).}
\end{footnotesize}
\end{minipage} 
\end{array}$
\end{figure*}

The \emph{forward--backward asymmetry} is
\begin{equation} 
\mbox{A}^{\! FB}_{b(\bar{b})} (s)
=
\sfrac{ \#_{b(\bar{b})}( \cos \theta_{b(\bar{b})} > 0 ) 
      - \#_{b(\bar{b})}( \cos \theta_{b(\bar{b})} < 0 )}
      { \#_{b(\bar{b})}( \cos \theta_{b(\bar{b})} > 0 ) 
      + \#_{b(\bar{b})}( \cos \theta_{b(\bar{b})} < 0 )}
\end{equation}
with $\cos \theta_{b(\bar{b)}} = \hat{p}_{e}.\hat{p}_{b(\bar{b})}$, 
where $\hat{p}_{x}$ is the direction of particle $x$. 

Since CP$(\mbox{A}^{\! FB}_{b}) = - \mbox{A}^{\! FB}_{\bar{b}}$, 
the \CPbar quantity is ${\mathcal O}^{FB} 
= \mbox{A}^{\! FB}_{b} + \mbox{A}^{\! FB}_{\bar{b}}$ (Figure~2)
\vspace{-6pt}
\be & & \hspace{-75pt}
= - 12 \alpha_{b}
  \left( \frac{2}{1 - \beta_{t}^{2}} 
  + \frac{1}{\beta_{t}} 
    \ln\left[ \frac{1 - \beta_{t}}{1 + \beta_{t}} \right] 
  \right)
  \mIm[ H_{2} ] / N_{t}
\\
\hspace{-45pt} \mbox{where} \hspace{45pt}
H_{2} 
&=& ( 1 - \Pol \bar{\Pol} ) H_{2}^{s} 
  + ( \Pol - \bar{\Pol} ) H_{2}^{a} 
\\ 
H_{2}^{s} 
&=& - c_{A} g_{A} h_{Z} \dgs
  + 2 c_{V} c_{A} g_{A} h_{Z}^{2} \dzs 
\\
H_{2}^{a}
&=& - c_{V} g_{A} h_{Z} \dgs
  + ( c_{V}^{2} + c_{A}^{2} ) g_{A} h_{Z}^{2} \dzs 
\; .
\ee

With the triple products ${\mathcal T}_{b(\bar{b})} =
\hat{p}_{e}.(\hat{p}_{t} \times \hat{p}_{b(\bar{b})})$ 
one can define the asymmetry $\mbox{A}^{\! {\mathcal T}}_{b(\bar{b})} = 
\frac{\#[{\mathcal T}_{b(\bar{b})}>0]-\#[{\mathcal T}_{b(\bar{b})}<0]}
     {\#[{\mathcal T}_{b(\bar{b})}>0]+\#[{\mathcal T}_{b(\bar{b})}<0]}$.
The \CPbar quantity is the difference 
\vspace{-6pt}
\be 
{\mathcal O}^{\mathcal T} 
= \mbox{A}^{\! {\mathcal T}}_{b} - \mbox{A}^{\! {\mathcal T}}_{\bar{b}}
= 3 \, \alpha_{b} \, \pi \, \sfrac{\sqrt{s}}{2 m_{t}} 
  \, \eRe[ D_{1} ] / N_{t}
\; ,
\ee
where $D_{1} = 
( 1 - \Pol \bar{\Pol} ) H_{1}^{a} + ( \Pol - \bar{\Pol} ) H_{1}^{s}$.

For the numercial analysis we take 
\mbox{$\sqrt{s} = 500$~GeV}, 
\mbox{$m_{\scriptscriptstyle W} = 80$~GeV}, 
$m_{t} = 175$~GeV, 
\mbox{$m_{b} = 5$~GeV}, 
\mbox{$\alpha_{s} = 0.1$}, \mbox{$\alpha_{em} = \sfrac{1}{123}$}, and
the following set of SUSY parameter values:
$M = 230$~GeV, $|\mu| = 250$~GeV, 
$m_{\tilde{t}_{1(2)}}=150 (400)$~GeV, 
$m_{\tilde{b}_{1(2)}}=270 (280)$~GeV,
$\tan\beta = 2$, $\varphi_{\mu}=\frac{4 \pi}{3}$,
$\theta_{\tilde{t}}=\frac{\pi}{9}$, 
$\varphi_{\tilde{t}}=\frac{\pi}{6}$, 
$\theta_{\tilde{b}}=\frac{\pi}{36}$, 
$\varphi_{\tilde{b}}=\frac{\pi}{3}$ 
and for the beam polarizations: 
$\Pol = 0.8$, $\bar{\Pol} = - 0.8$.

In Figure~1 we show $\mbox{R}(s)$ and in Figure~2 we show 
${\mathcal O}^{FB}$ for the set of parameters given above. 
The \CPbar ratios $\mbox{R}(s)$ and ${\mathcal O}^{FB}$ depending on
$\mIm\dgs$ and $\mIm\dzs$, show spikes (see Fig.~1). They are due to 
threshhold effects in the fermion pair production in the loop. 
These spikes only appear in the chargino and neutralino contributions. 

\section*{Acknowledgements}
We thank J.W.F. Valle and all the organizers for the invitation to 
the interesting workshop and the kind hospitality in Valencia. 
We thank Helmut Eberl, Sabine Kraml and Stefano Rigolin for the 
helpful discussions. E.C.'s work has been 
supported by the Bulgarian National Science Foundation, Grant Ph--510. 
This work was also supported by the 'Fonds zur F\"orderung der 
wissenschaftlichen Forschung' of Austria, project no. P10843--PHY. 

\section*{References}
\begin{flushleft}

\end{flushleft}


\begin{thebibliography}{99}
\bibitem{bcgm:1}
  see references [1] and [2] of~\cite{bcgm}
 
\bibitem{we} 
  A.~Bartl, E.~Christova, W.~Majerotto, 
  \emph{ Nucl.Phys.} \textbf{B460} (1996) 235; erratum 
  \emph{ Nucl.Phys.} \textbf{B465} (1996) 365;

\bibitem{Kane}
  H.~E.~Haber, G.~L.~Kane,
  \emph{ Phys.Rep.} \textbf{177} (1985) 75

\bibitem{Dugan} 
  M.~Dugan, B.~Grinshtein, L.~Hall, 
  \emph{ Nucl.Phys.} \textbf{B255} (1985) 413; \\
  W.~Bernreuther, M.~Suzuki 
  \emph{ Rev.Mod.Phys.} \textbf{63} (1991) 313; \\
  W.~Hollik, J.I.~Illana, S.~Rigolin, D.~St\"ockinger, 
  \textbf{ hep--ph} / 9711322

\bibitem{bcgm}
  A.~Bartl, E.~Christova, T.~Gajdosik, W.~Majerotto, HEPHY-PUB 
  669/97, UWThPh-1997-13, \textbf{hep--ph} / 9705245, 
  to be published in \emph{ Nucl.Phys.} \textbf{B}
 
\bibitem{Osh1} 
  Y.~Kizukuri and N.~Oshimo, 
  \emph{ Phys.Rev.} \textbf{D45} (1992) 1806; 
  \emph{ Phys.Rev.} \textbf{D46} (1992) 3025

\bibitem{Garisto}
  R.~Garisto, J.~D.~Wells, 
  \emph{ Phys.Rev.} \textbf{D55} (1997) 1611

\bibitem{SingW}
  J.~M.~Ortega,
  {\em Matrix Theory} (Plenum Press, New York 1987)

\bibitem{PaVe}
  G.~Passarino and M.~Veltman, 
  \emph{ Nucl.Phys.} \textbf{B160} (1979) 151; $\:$
  A.~Denner,
  \emph{ Fortschritte der Physik} \textbf{41} (1993) 4, 307

\bibitem{cd}
  E.~Christova, D.~Draganov, 
  \textbf{ hep--ph} / 9710225

\end{thebibliography}
\end{document}